\documentclass[twocolumn,showpacs,showkeys,nofootinbib]{revtex4}
\usepackage{graphicx}
\usepackage{soul}
\usepackage{color}
\usepackage{amsmath}
\usepackage{amssymb}
\usepackage{pgf}
\usepackage{subfigure}
\newcommand{\be}{\begin{eqnarray}}
	\newcommand{\bel}{\begin{equation}\label}
		\newcommand{\ee}{\end{equation}}
	\def\dg{\dagger}
	\newcommand{\barl}{\begin{eqnarray}\label}
		\newcommand{\ear}{\end{eqnarray}}
	\newcommand{\non}{\nonumber}
	
\begin{document}
		\title{Qubit -- Photon Bound States in Superconducting Metamaterials}
		\author{M. Peji\'c$^1$, \v Z. Pr\v zulj$^1$, D. Chevizovich$^1$, 
			N. Lazarides$^{2,3,4}$, G. P. Tsironis$^{2,3,4}$, Z. Ivi\'c$^{1,2,4}$}
		\affiliation{$^{1}$ Laboratory for Theoretical and Condensed Matter Physics, "VIN\v CA" Institute of Nuclear Sciences,
			National Institute of the Republic of Serbia, University of Belgrade 
			P.O.Box 522, 11001 Belgrade, Serbia \\ 
			$^{2}$Institute of Theoretical and Computational Physics, Department of Physics, 
			University of Crete, P.O. Box 2208, Heraklion, 71003, Greece \\
			$^{3}$Institute of Electronic Structure and Laser,
			Foundation for Research and Technology--Hellas, P.O. Box 1527, 71110 
			Heraklion, Greece \\
			$^{4}$National University of Science and Technology MISiS, Leninsky Prospekt 4, 
			Moscow, 119049, Russia
		}
		\date{\today}
		\begin{abstract}
			We study quantum features of electromagnetic radiation propagating in the one--dimensional superconducting quantum metamaterial comprised of an infinite chain of charge qubits placed within two--stripe massive superconducting resonators. The Quantum--mechanical model is derived assuming weak fields and that, at low temperatures, each qubit is either unoccupied ($N=0$) or occupied by a single Cooper pair ($N=1$). We demonstrate the emergence of two bands of single--photon--qubit-bound states with the energies lying outside the photon continuum: highly above and slightly below it. The higher energy band slowly varies with the qubit--photon center of mass quasi-momentum. It becomes practically flat provided that electromagnetic energy is far below the Josephson one, while the latter is small compared to the charging one. The dispersion of the lower band is practically identical to that of free photons. The emergence of bound states may cause radiation trapping indicating its application for the control of photon transport in superconducting qubit--based artificial media. 
		\end{abstract}
		\maketitle
		\section{Introduction}
		During the last two decades, there has been great progress in the design of quantum devices based on waveguide structures. The latter is comprised of quantum emitters of natural or artificial atoms (qubits) that are coupled with the one--dimensional optical channel. Numerous applications in quantum simulation, quantum information processing, and communication have already been discussed in the literature\cite{qed1, qed2, qed3, qed4, QIT1, QIT2, QIT3, QIT4, QIT5}. Despite all efforts, the level of control and preservation of quantum coherence achieved in these media are still beyond the requirements for a practical realization of operable quantum information devices. This difficulty could be overcome by a better understanding of the nature of the interaction between matter and radiation. Accordingly, intensive studies of light--matter interaction in waveguide structures are necessary. 
		Superconducting metamaterials (SCQMMs) are man-made material units that are both interesting and quite promising for the ultimate fabrication of quantum devices \cite{zag, qmm1, qmm2, qmm3, Rakhmanov2008, Shvetsov2013, Asai2015, Asai2018}. These engineered media, comprised of periodically arranged artificial atoms that form superconducting quantum bits (SCQ) while interacting with EM fields inside one--dimensional transmission lines. Owing to spatial confinement, tunability of the SCQ parameters, and the ability to tailor photon dispersion relation by their particular setup, SCQMMs may be conveniently engineered to provide tunable ``atom'' -- field interaction that can reach regimes ranging from weak to ultra strong coupling. This is of particular interest in the case of qubit interaction with quantized radiation fields when strong qubit--photon coupling leads to effective photon--photon and qubit--qubit correlations. The latter allows for the emergence of novel interference effects with possible practical applications. For example, photons may exhibit a nontrivial dispersion relation such as band edges and band gaps. In this way, QMM may be viewed as a photonic crystal \cite{qmm0}. This enriches their potential for practical applications and provides novel means for devising comprehensive studies of practical and fundamental aspects of the artificial atom--field interaction. 
		Investigations of the emergence of an atom (or other emitter, qubit in particular)--photon bound states \cite{aphbs1, aphbs2, aphbs3, aphbs4, aphbs5, aphbs6, aphbs7, aphbs0, prohib, aphbs8, aphbs9, aphbs10,aphbs11} are of particular importance due to their consequences for radiation propagation \cite{prohib,ilja, prohib2, prohib3}, preservation of quantum coherence and entanglement \cite{ilja,ent1,aphbs11}. For example, the prohibition of the free propagation of radiation could be attributed to the formation of these bound states. Their creation within the continuum potentially can be used for the storage of quantum information \cite{prohib,prohib2,prohib3} and construction of photon memory devices \cite{aphbs3}. 
		On the other hand, the recent discovery of topological excitations in SCQMMs implies that, by the engineering of topologically nontrivial QMM, it would be possible to tackle unavoidable structural irregularities in SCQMMs. This is since the creation of photon bound states provides the preservation of quantum coherence for times large enough to perform quantum information processing \cite{ilja}. A further important possible application is the exploitation of qubit--photon bound states as a mechanism of the entanglement preservation in quantum information processing \cite{ent1,aphbs11}.
		In this paper, we study the qubit--photon bound states emerging through the interaction of an EM field propagating through SCQMM, consisting of massive, two stripe-superconducting resonator filled with a large number ($\mathcal{N}\gg 1$) of Cooper pair box (CPB) or charge qubits. Such, essentially three--dimensional (3D) structure, substantially differs from the most common realization of SCQ based waveguiding structures \cite{qed1} and SCQMM setups \cite{qmm1} in which point--like SCQs are built-in \textit{coplanar} resonators. In such two--dimensional (2D) architectures, qubit--photon interaction is described within Jaynes (Tavis)--Cummings \cite{qed1}, Dicke model Hamiltonians \cite{aphbs8,aphbs9}. 
		A realistic theoretical model for the proposed setup is derived in the next section in terms of classical variables is, while its quantization is performed in the third section. It substantially differs from those used in the studies of the consequences of ``atom''--light interaction in the engineered media so far \cite{aphbs8, aphbs9}, i.e. modified Dicke Hamiltonian. In particular, while the pure photon part is practically identical to those encountered so far, the qubit -- photon interaction is quite different and contains two terms that may be attributed to attractive and repulsive interaction, whose competition determines the character and existence qubit--photon bound states. 
		The paper is organized as follows: Description of the model and classical Hamiltonian are introduced in the second section. The quantization procedure is given in the third section. Two--particle Schr\"odinger equation and its solutions are discussed in Section 4. Results and conclusions are summarized in the fifth section. Details of the mathematical derivation are given in the Appendices.
		\section{Experimental setup: the proposal}
\begin{figure}[h]
			\begin{center}
				\includegraphics[height=7cm]{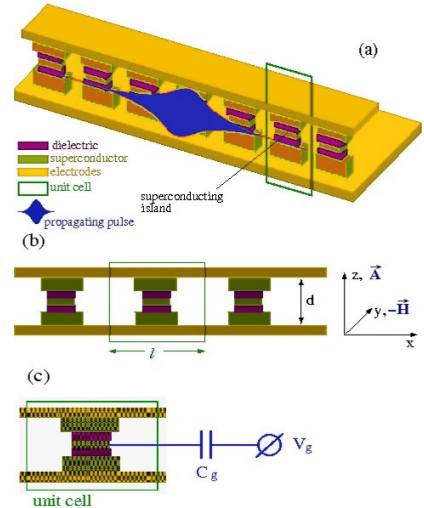}
				\vskip -0.5 cm
				\caption{Illustration of the proposed setup of SCQMM: (a) A chain of Cooper pair box qubits inside the two--stripe transmission line. Each unit cell contains a tiny superconducting island connected with TL banks through two Josephson junctions, for the regions of the dielectric layers (blue). The propagating electromagnetic vector potential pulse is also shown schematically out of scale.\\
					(b) The side view of the SCQMM. The magnetic field penetrates through free space between the islands.\\
					(c) A unit cell of the SCQMM showing the control circuitry of the charge qubit, consisting of a gate potential $V_g$ applied to it through the gate capacitor $C_g$.}
				\label{fig1}
			\end{center}
		\end{figure}

		We investigate the non--classical properties of electromagnetic radiation propagating along the SCQMM in a construction visualized in Fig. (\ref{fig1}). It is made of an infinite ($\mathcal{N}\rightarrow \infty$), one--dimensional (1D) periodic SCQ array, with period $\ell$, placed in a transmission line (TL) consisting of two infinite bulk superconductors separated by a distance $d$; we consider $d$ being of the same order of magnitude as $\ell$ \cite{Rakhmanov2008, Shvetsov2013, Asai2015, Asai2018} (Figs. 1a and b). For simplicity, we took that the thickness of superconducting strips is $\ell$. Each SCQ is a tiny superconducting island connected to each bank of the TL through a Josephson junction (JJ). The control circuitry for each SCQ (Fig. 1c), consists of a gate voltage source $V_g$ coupled to it through a gate capacitor $C_g$ and allows for local control of the SCQMM by altering independently the state of each SCQ \cite{zag}. The SCQs exploit the nonlinearity of the Josephson effect and the large charging energy resulting from nanofabrication to create artificial mesoscopic two-level systems.
		\subsection{Classical model Hamiltonian}
		In order to set up the problem, we first derive the classical model and subsequently perform its quantization. We assume that an electromagnetic (EM) wave with vector potential $\vec{A}=A_z (x,t) \hat{z}$ propagates along with the superconducting TL. The direction of propagation is parallel to the superconducting electrodes and while $\vec{A}$ is perpendicular to the direction of the EM wave propagation.
		Let us now derive the model Hamiltonian of the system under the consideration. First, we recall, that each CPB qubit is comprised of \textit{double barrier} JJ (DBJJ), i.e. two JJs connected in a series. Similar DBJJ has been widely studied and used in different contexts \cite{zag, tinkh, JJrow, JJrow1, zag1}.
		Hamiltonian of DBJJ may be obtained employing the straightforward extension of the Feynman semi--classical approach \cite{feynm} in which the dynamics of the single JJ is described within the simple two--level model: wave function of the CP condensate (Ginzburg--Landau -- GL order parameter) on each side of JJ is represented as $\varPsi_{p=1,2}=\sqrt{n_p}e^{i \phi_p}$, while the Cooper pair tunneling is accounted through the phenomenological coupling term. In such a way, single JJ dynamics may be described employing the nonlinear model Hamiltonian of the single variable: \textit{Josephson phase} representing the difference between the phases of GL order parameter in particular superconductors.
		\begin{figure}[h!]
			\begin{center}
				\hspace{-1cm}
				\includegraphics[width=0.55\textwidth]{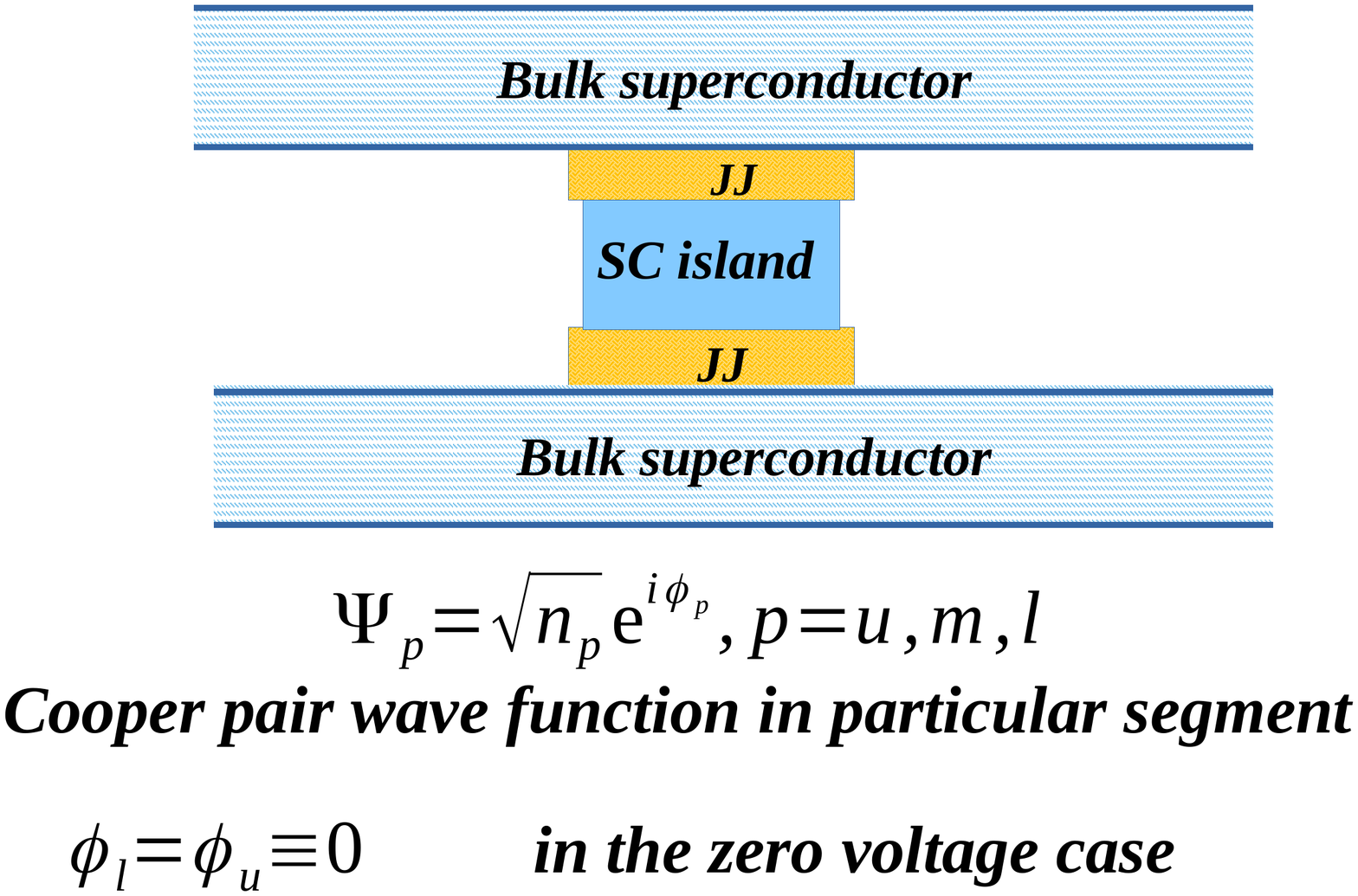}
				\vspace{-1.5cm}
				\caption{Simplified illustration of the CPB qubit 
					consisting of two Josephson junctions connected in series.}
				\label{fig2}
			\end{center}
		\end{figure}
		In the case of DBJJ, three superconducting segments separated by two JJs (\ref{fig2}), wave function in particular segment, upper, middle and lower, may be written as: $\Psi_p(t)=\sqrt{n_p}e^{i\phi_p(t)},\;\; p=u,m,l$, while the tunneling between them now is simply: $$H_t=-V\left(\Psi^*_u\Psi_m+\Psi^*_m\Psi_l+c.c\right)$$
		where $V$ represents phenomenological parameter so called Josephson constant \cite{feynm}. 
		Substitution of the CP wave function as given above and, in analogy with \cite{feynm}, assuming that CP numbers in each segment are almost the same and equal to $n_0$, we found that the Hamiltonian of double JJ system is the sum of Hamiltonians of two independent JJs:
		\begin{eqnarray}\label{single1}
			H= -E_c\sum_{i=l,u}\frac{\partial^2}{\partial \varphi_{i}^2}-E_{J}\sum_{i=l,u}\cos{\varphi_{i}},
		\end{eqnarray} 
		\noindent Here $\varphi_{u,(l)}$ denote the Josephson phase differences at lower (upper) junction. In our model we restrict ourselves to \textit{zero voltage} case \cite{Asai2015} when $\phi_l=\phi_u\equiv 0$ so that these phases read: $\varphi_{u}=\phi_u-\phi_m\equiv -\phi_m$ and $\varphi_{l}=\phi_l-\phi_m\equiv -\phi_m$.
		The energy parameters $E_c=\frac{2e^2}{C_J}$ and $E_J=\frac{\Phi_0 I_C}{2\pi c}$, $\Phi_0=\frac{h c}{2 e}$, $I_c$, $C_J$, and $c$ are the junction charging energy, so called Josephson energy, flux quantum, critical current, junction capacity, and speed of light, respectively.
		In the presence of EM field Josephson phase difference $\varphi_{i}$ acquires the gauge term and reads:
		\bel{gauge} \varphi_u(t)-\varphi_l(t)=-\phi_m\pm \frac{4\pi}{\Phi_0}\int_{1}^{2}\vec A(\vec r)\cdot d\vec l. \ee
		Generalizing (\ref{single1}) to whole qubit lattice and accounting for the energy of EM field inside the SCQMM ($H_{em}=
		\frac{1}{8\pi}\int\left( E^2_n(\vec r)
		+B^2_n(\vec r) \right) d^3r$) we derived a total model Hamiltonian: 
		\begin{eqnarray}
			\label{tot}
			\nonumber 
			H=&&\sum_n\bigg[\frac{2\hbar^2}{E_c}\dot{\phi}^2_n -2E_J\cos\phi_n\cos\alpha_n+\\ 
			&&\frac{2\hbar^2}{E_c}\dot\alpha^2_n+ \dot\alpha^2_n +E_{em}(\alpha_{n+1}-\alpha_n)^2\bigg].
		\end{eqnarray}
		Here we have introduced the dimensionless amplitude of the vector potential $\alpha_n=\frac{2\pi d}{\Phi_0}A_n$.
		To facilitate practical calculation the charging term has been redefined using $\frac{d \phi_n(t)}{d t}\equiv \dot{\phi_n}=-i\frac{2e^2}{\hbar C_J}\frac{\partial}{\partial \phi_n}$, while the gauge term in Josephson phase difference has been approximated as $\int_{1}^{2}\vec A(\vec r) \cdot d\vec l\equiv \frac{2\pi d}{\Phi_0}A_n$. The identity holds in the present setup where it was assumed that the separation $d$ between the superconducting stripes and the period $\ell$, i.e. the center--to--center distance between qubits, are of the same order of magnitude and much smaller than the wavelength of EM radiation; this fact enables us to neglect the variation of the vector potential within each cell. As a result, the integration in Eq. (\ref{gauge}) is trivial (see for example \cite{Rakhmanov2008, Shvetsov2013, Asai2015}).
		Finally, in evaluating EM energy integration is taken over the entire unit cell. Thus, in accordance with the approximation adopted above, we neglect spatial variation of electric and magnetic field within the unit cell, so that the energy of EM field in particular unit cell was approximated as:
		\begin{eqnarray}\label{emfapp}
			\non && H_{em}\approx \frac{V}{8\pi}\left( E^2_n+B^2_n\right), \\
			&& V=\ell^2d, \; \; - \mathrm{volyme\;\; of\; the\;\; unit\; cell. }
		\end{eqnarray}
		For simplicity, we took that the width of superconducting stripes is equal to inter--qubit distance: $\ell$. 
		Following \cite{Rakhmanov2008} and \cite{Asai2015} we have neglected the contribution of the electric field, while the fraction that originates from the magnetic field was accounted for through the discretization procedure introduced in \cite{Rakhmanov2008, Shvetsov2013, Asai2015}:
		\begin{equation}\label{mf}
			B(x,t)=\frac{\partial A(x,t)}{\partial x}\rightarrow \frac{A^z_{n+1}-A^z_n}{\ell}.
		\end{equation} 
		Here $E_{em}=\frac{1}{8\pi \ell d}\bigg(\frac{\Phi_0}{2\pi}\bigg)^2$, is the so called \textit{electromagnetic energy} introduced in \cite{Rakhmanov2008}, determining the speed of "light" in the qubit chain, which, in dimensionless units, reads $\beta=\sqrt{E_{em}/E_J}$. It, together with the ratio $\gamma=\frac{E_C}{E_J}$ represents the main quantitative characteristic of CPB qubits, their derivatives, transmon for example, and networks made of them. 
		\section{Quantization and two--level approximation}
		The quantum--mechanical versus a (semi)classical description of the qubit--EM field coupled systems still has certain controversies \cite{contro}. Nevertheless, at low temperatures, a fully quantum treatment is justified, while the dissipation is negligible. Under these conditions, the quantum state of an island is determined by the number of extra Cooper pairs on them. In addition, EM radiation exhibits quantum features for weak (small amplitude) EM fields when their modes are populated with just a few photons, one or two, per wavelength \cite{fist2}. 
		At this stage, we must note that the tunneling of the single CP between the banks and island does not affect the state of the former which contains a large number of CPs so that the deficiency or the excess of the single CPs has no particular significance.
		Formally we quantize our model by introducing the photon creation and annihilation operators in real (direct) space and Josephson phase and Cooper pair number operator in Cooper pair number basis. In such a way, through a few intermediate steps, described in Appendix 1, the classical Hamiltonian Eq. (\ref{tot}) can be approximated by the quantum one describing the interaction of a collection of the two--level systems and the quantized multimode electromagnetic field.
		\subsection{EM field}
		In the quantum regime the electromagnetic field is weak, i.e., the dimensionless amplitude of its vector is small and can be treated as quantum fluctuation, i.e. $\alpha_n\rightarrow\hat \alpha_n\ll 1$. This enables us to expand $\cos\hat\alpha_n\approx 1-\frac{\hat\alpha^2_n}{2}$. Next, we quantize the EM field in two steps: first we define the generalized momentum $P_n=\frac{2\hbar^2}{E_c}\dot{\alpha_n}$ canonically conjugated to $\alpha_n$. Subsequently we treat photon variables as operators $\alpha_n\rightarrow\hat \alpha_n$, $P_n\rightarrow \hat P_n$ satisfying the commutation relation $[\hat\alpha_n,\hat P_m]=i\hbar \delta_{m,n}$. It holds for the transformation Eq. (\ref{quantiz1}) through which we introduce photon creation and annihilation operators in real (direct) space:
		\begin{eqnarray}\label{quantiz1}
			\hat \alpha_n=
			\frac{1}{2}\sqrt{\frac{E_C}{\hbar\omega}}\left(a_n+a^{\dg}_{n}\right),\;\;
			\hat P_n={i\hbar}\sqrt{\frac{\hbar\omega}{E_c}}\left(a^{\dg}_{n}-a_n\right).
		\end{eqnarray}
		\subsection{Qubit subsystem}
		Similarly, in quantization of the CPB qubit subsystem we introduce the pair canonically conjugated variables (operators): the \textit{phase} $\phi\rightarrow \hat \phi$ and Cooper pair number operator $\hat N=-i\frac{\partial}{\partial \hat\phi_n}$, $[\phi_n, \hat{N_n}]=i$. Then we rewrite Eq. (\ref{tot}) in the Cooper pair number basis $|N\rangle$, using the correspondence: $\hat N=-i\frac{\partial}{\partial \phi_n}$ and noticing that $e^{\pm i\hat \phi_n }|N\rangle = |N \pm 1\rangle$.Next, in the obtained Hamiltonian we exploit the fact that in \textit{charge and transmon} regime only a few lowest levels are relevant and we may restrict ourselves to the reduced state space in which the single island can be unoccupied ($N=0$) or occupied by single Cooper pair ($N=1$). The resulting Hamiltonian is nondiagonal in reduced number basis ${|0\rangle,\; |1\rangle}$, and in the next step we diagonalize the free qubit part by means of transition to energy eigenbasis (${|e\rangle}-\mathrm{excited\; state},\; |g\rangle-\mathrm{ground\; state}$) performing the norm preserving unitary transformation Eq. (\ref{UT}). Finally, after neglecting the photon number non-preserving terms, i.e., those $\sim a^2_n$ and $\sim a^{\dg 2}_n $, we obtain the quantized model Hamiltonian:
		\begin{eqnarray}\label{qb}
			\non H=&&\Delta\sum_n|e\rangle_n\langle e|+\\
			&&\hbar\omega\sum_n a^{\dg}_n a_n-J\sum_n a^{\dg}_n(a_{n+1}+a_{n-1})+\\ \non &&\sum_n\bigg[B(|e\rangle_n\langle g|+|g\rangle_n\langle e|) -A|e\rangle_n\langle e|\bigg]a^{\dg}_n a_n
		\end{eqnarray}
		Here the first term represents the Hamiltonian of the qubit subsystem with level splitting between the excited and ground-state $\Delta=2\epsilon$ ($\epsilon=\sqrt{E^2_J+E^2_C}$). It is represented here in terms of the operator $|e\rangle\langle e |$ to emphasize that initially system is prepared so that all qubits are excited. Such "atoms" are usually called \textit{emitters}.
		In the pure photon Hamiltonian, the two terms in the second line, correspond to typical boson \textit{tight binding} model describing photon hopping between neighboring qubits. Parameters $\omega$ and $J$ stay for the photon frequency and the photon inter--qubit tunneling amplitude, respectively: 
		\begin{eqnarray}
			\label{param1}
			\hbar\omega=\sqrt{2E_{em}E_C+\frac{E_CE^2_J}{2E}}, 
			\qquad 
			J=\frac{E_{em}E_C}{2\hbar\omega}. 
		\end{eqnarray}
		Considering the noninteracting case, pure photon, and qubit system, the present model is analogous to those appearing frequently in a theoretical description of charge and energy transfer in various contexts. Recent application concerns the photonic bandgap materials where it addresses the photon hopping motion in \textit{coupled resonator(cavities) waveguides} \cite{aphbs9,aphbs10}. Quantum metamaterials built of such structures with embedded tunable quantum emitters, i.e., qubits, opened a new perspective for further development of novel, quantum, technological devices, and for studies of nonclassical features of light \cite{fist2}. 
		Finally, the last term is related to the qubit -- photon interaction. It possesses two components: the attractive one, measured by the parameter $A$, and repulsive $\sim B$.
		\begin{eqnarray}\label{param2}
			A=\frac{E^2_JE_c}{4\hbar\omega \epsilon},\;\; B=\frac{E_JE^2_c}{8\hbar\omega \epsilon}.
		\end{eqnarray}
		For the convenience we rewrite the interaction Hamiltonian in terms of "atomic" (pseudo--spin) operators ($\sigma^{\dg,-,z}$):
		\begin{equation}\label{atomic}
			H_i=\sum_n [B(\sigma^{\dg}_n+\sigma^{-}_n)-A\sigma^{\dg}_n\sigma^{-}]a^{\dg}_na_n.
		\end{equation}
		The operators in the attractive interaction term may be rearranged as follows: $\sigma^{\dg}\sigma^-a^{\dg}a\equiv \sigma^{\dg}a\sigma^-a^{\dg}- \sigma^{\dg}\sigma^-$. Thus, it may be understood to originate on account the simultaneous excitation ($\sigma^{\dg}a$) and de--excitation ($\sigma^{-}a^{\dg}$) of the $n$--th qubit by an absorption and emission of the single photon. On the other hand, repulsive interaction comes from the photon scattering by qubits resulting in their excitation ($|e\rangle\langle g|$) and de--excitation ($|g\rangle\langle e|$).
		These mechanisms differ substantially from those accounted for within the Dicke and Jaynes--Cummings models coming from the excitation qubit, atom in general, by the absorption of the single photon ($|e\rangle\langle g| a$) and vice versa: qubit deexcitation by the emission of the single photon. 
		In coplanar geometry setups \cite{qed1,qed2} a qubit--photon interaction substantially differs from the present one, and, in the rotating wave approximation, reads:
		$$H_{JC}=g\sum_n \sigma^{\dg}_n a_n+\sigma^{-}a^{\dg}_n$$. 
		Note that here we can not distinguish whether the interaction is attractive or repulsive. This becomes possible only after deriving the exigent--value equation, counterpart equation (\ref{eigen1}) from the next paragraph, based on the sign of effective interaction parameter. 
		So far, the interaction resulting from the set--up proposed here was not encountered in the studies of the light interacting neither with natural nor artificial media. Nevertheless, formally very similar models may appear in solids, magnetic semiconductors \cite{sd}, when a single electron creates micro--ferromagnetic domains flipping the spins of neighboring ions, while the interaction Hamiltonian is given in terms of the $s-d(f)$ being very similar to (\ref{atomic}.)
		We also point out that, in the present setup, the waveguide is the chain of unit cells (sketched at Fig. 1b) each of which contains a single qubit ("atom") and may be viewed as an optical resonator. That is, our waveguide is the set of the large number ($\mathcal{N}\gg 1$) of coupled resonators (unit cells) with one "atom" per "cavity", which imply translational invariance of the system. Nevertheless, most often, the waveguide is the set "resonators" designed independently of "atoms". In these structures "atoms" are arranged arbitrarily, depending on the particular application--research subject. Various settings are possible and a particular waveguide may be populated by a few ($\mathcal{N}$) "atoms", with one or more "atoms" per cavity \cite{aphbs3, aphbs4, aphbs5, aphbs6, aphbs7, aphbs0, prohib, aphbs8, aphbs9, aphbs10, aphbs11}. 
		One more distinction must be made in comparison with related systems. In that respect we refer to quantum metamaterial designed of coplanar, mostly superconducting, resonator waveguide and several embedded qubits \cite{qed1, qed2, qed3, qed4}, where the qubits are linearly\footnote[2]{The interaction is of the first order in field amplitude and contains only the terms linear in photon operators.} coupled to the resonator modes. 
		\section{qubit photon bound states}
		\subsection{Vector of state and Schr\"odinger equation}
		The wave function which diagonalizes Hamiltonian Eq. (\ref{qb}) has a form of a 
		single photon dressed qubit (atom) state:
		\begin{eqnarray}\label{psi}
			\non |\Psi\rangle= 
			&&\sum_m u_m a^{\dg}_m|0\rangle|g\rangle +\sum_{m,n}\Psi_{m,n} 
			\sigma^{\dg}_n a^{\dg}_m |0\rangle|g\rangle_m, \\ 
			&&\sigma^{\dg}_n=|e\rangle_n\langle g|,\;\; \Psi_{m,n}=\Psi_{n,m}.
		\end{eqnarray}
		Here the probability amplitudes satisfy the normalization condition
		\begin{equation}
			\label{norm}
			\sum_m |u_m|^2+\sum_{m,n}|\Psi_{m,n}|^2=1. 
		\end{equation}
		The first term in state Eq. (\ref{qb}) corresponds to the case when a single photon is excited in site $m$ with probability amplitude $u_m$, while the qubit remains in its ground state. The second term of a vector of state Eq. (\ref{qb}) corresponds to synchronized excitation of $n$--th qubit and photon at site $m$. The symmetry property $\Psi_{m,n}=\Psi_{n,m}$ reflects the translational invariance of chain: solutions must remain invariant when photon and qubit excitation exchange position simultaneous excitation of qubit at site $m$ and photon at $n$--th site. 
		Owing to orthogonality of $\langle g|\langle 0|a_m$ and $\langle g|\langle 0|\sigma^{-}_m a_n$ and $|\Psi\rangle$ we may project Schr\"odinger equation $H|\Psi>=E|\Psi>$ onto $\sigma^{\dg}_m a^{\dg}_n|g\rangle|0\rangle$ and $a^{\dg}_m |g\rangle|0\rangle$. In this way we obtain a system of coupled equations for the amplitudes $\Psi_{m,n}$ and $u_m$:
		\barl{SE}
		\non &&(\mathcal{E}-\Delta)\Psi_{m,n} 
		+\frac{J}{2}\bigg(\Psi_{m,n+1}+\Psi_{m,n-1}
		+\left\lbrace m \rightleftarrows n \right\rbrace \bigg)=\\ 
		&&\non -A\Psi_{m,n}\delta_{m,n}+B u_m\delta_{m,n},\\
		&& \mathcal{E}u_m+J(u_{m+1}+u_{m-1})=B\Psi_{n,n}.
		\ear
		We will solve it by employing Fourier transform. Owing to the translational invariance we pick:
		\begin{equation}
			\label{ft}
			\Psi_{m,n}=\frac{1}{\sqrt{\mathcal{N}}}e^{i\frac{K(m+n)}{2}\ell}\Phi_{m-n},\;\; 
			u_m=\frac{1}{\sqrt{\mathcal{N}}}\sum_k u_k e^{ikm\ell}. 
		\end{equation} 
		In this way, the second equation in Eq. (\ref{SE}) attains a simple form and may be readily solved for $u_m$, which then may be eliminated from the first one. In the resulting equation we employ the translational invariance and took $m-n=l$; next we perform Fourier transform $\Phi_l=\frac{1}{\mathcal{N}^{1/2}}\sum_q \Phi_q e^{iql\ell}$. This finally yields:
		\begin{eqnarray}
			\label{basic}
			&&\non [\mathcal{E}-\Delta+2J\cos(Kd/2) \cos q] \Phi_q=\\ 
			&&\bigg[-A+\frac{B^2}{(\mathcal{E}+ 2J\cos K)} \bigg]
			\left(\frac{1}{\mathcal{N}}\sum_q\Phi_q\right),
		\end{eqnarray}
		here $K$ and $q$ stand for center of mass and relative qubit--photon quasi--momenta, while $\mathcal{E}=E-\hbar\omega$.
		On the basis of this equation it is easy to find relation for eigenvalues: we first find $\Phi_q=...$, then we multiply both sides of the last equation with $1/\mathcal{N}$ and then sum up both sides over $q$. This results in:
		\begin{equation}
			\label{eigen0} 
			1=\frac{1}{\mathcal{N}}
			\sum_q\frac{1}{(\varepsilon-\delta+ \cos (K\ell/2)\cos q\ell)}\bigg[-a
			+\frac{b^2}{(\varepsilon + \cos K\ell)} \bigg].
		\end{equation}
		Bound state solutions, if any exist, must lie outside the free state continuum appearing in the absence of qubit--photon interaction. In that case Eq. (\ref{basic}) has solution
		\begin{equation}
			\varepsilon(q,K)=\delta - \cos q\ell\cos \frac{K\ell}{2},
		\end{equation}
		so that the bound state energy must lie either below the lower energy bound $$ \delta-|\cos(K\ell)/2|,$$or above the higher one $$ \delta+|\cos(K\ell)/2|.$$
		\subsection{Eigenvalue equation}
		The summation over $q$ may be performed in accordance with the rule: $\frac{1}{\mathcal{N}}\sum_q<....>=\frac{1}{2\pi \ell}\int_{-\pi/\ell}^{\pi/\ell}d q<...>$. This, provided that $|\varepsilon-\delta|>1$, yields the self--consistent equations for energy eigenvalues:
		\begin{eqnarray}\label{eigen1}
			\non 1=a'(K) \frac{sgn{(\varepsilon-\delta)}}{\sqrt{(\varepsilon-\delta)^2-\cos^2 K\ell/2}},\\
			a'(K)=-a+b^2\frac{1}{\varepsilon+\cos K\ell},
		\end{eqnarray}
		where $a=A/2J$, $b=B/2J$, $\delta=\Delta/2J$ and $\varepsilon=\mathcal{E}/2J$, stand for the normalized coefficients. For further convenience we express Eq. (\ref{eigen1}) in terms of just two parameters, $\beta$ and $\gamma$, which fully characterize proposed system:
		\begin{eqnarray}
			\label{norpar}
			\non a=\frac{1}{4\beta^2}\frac{1}{\sqrt{1+\gamma^2}},\; 
			b=\frac{\gamma a}{2},\\ \frac{\hbar\omega}{2J}=2+\frac{1}{2\beta^2\sqrt{1+\gamma^2}},\\
			\non \delta=2\sqrt{2\frac{(1+\gamma^2)}{\gamma\beta^2}+\frac{\sqrt{1+\gamma^2}}{2\gamma\beta^4}}.
		\end{eqnarray}
		Eigen--equation (\ref{eigen1}) is a nonlinear (in $\varepsilon(K)$) transcendental equation and can not be solved analytically. Nevertheless, its nonlinearity implies that it may have multiple solutions. That is, qubit--photon bound states if any exist, should exhibit multi--band structure. 
		To facilitate practical calculations, to examine the possible appearance of multi--band structure of the qubit--photon spectra, finally, to compare present analysis with the related preceding ones \cite{bs1,bs2} we rewrite Eq. (\ref{eigen1}) in the self-consistent form
		\begin{equation}\label{selfc}
			\varepsilon(K) -\delta=\pm \sqrt{a'^2(K)+\cos^2 K/2},
		\end{equation}
		in which, on the right hand side, $\varepsilon(K)$ appears implicitly through the $a'(K)$ in accordance with Eq. (\ref{eigen1}). 
		This "solution" recalls much the exact one in the limit $a'(K)\rightarrow a$, appearing frequently in different contexts. Examples are numerous, and, despite different physical backgrounds, formally identical solutions, may be found in many cases such as \textit{bound states of two photons, phonons, excitons} \cite{bs1}. In addition, the problem of the bound state of an impurity atom and its vibrational or magnetic environment \cite{bs1}, within the simplest models, also reduce to this elementary solutions. 
		\subsection{Existence of solutions}
		Solubility of Eq. (\ref{eigen1}) requires non--negativity of its right hand side, thus, for $\varepsilon - \delta<0$ ($\varepsilon - \delta >0$), eigen--energy solutions exist provided that $a'(K)<0$ ($a'(K)>0$). Accordingly, signs ($+$ or $-$) in Eq. (\ref{selfc}) stand for $\varepsilon - \delta<0$ and $\varepsilon - \delta>0$, respectively. Also, throughout the paper, we may call $a'(K)$ the \textit{effective} qubit--photon interaction strength. The term ``effective'' is used here to emphasize the self--consistency of (\ref{selfc}), and to point to its \textit{formal equivalence} with the exact ones appearing when $a'(K)\rightarrow a$.
		To find $\varepsilon(K)$ we have performed the numerical calculation focusing ourselves to the case $\varepsilon - \delta<0$ when an effective qubit--photon interaction is attractive. An opposite case was not considered since our numerical calculations have shown that the solutions of the eigenvalue problem exist for unrealistic values of system parameters. For example for $\gamma \sim 100$.
		\subsection{Solutions: analytical considerations}
		Before presentation of the results our numerical calculations we perform some auxiliary analytic analysis evaluating explicitly eigen--energies at band edges: $\varepsilon(\pm \pi)\equiv \varepsilon(\pi)$. In that limit (\ref{selfc}) become:
		\begin{equation}
			\label{edge}
			\varepsilon(\pi) -\delta
			=\pm a\left(1-\frac{a(\frac{\gamma}{2})^2}{\varepsilon(\pi)-1}\right).
		\end{equation}
		The signs $(+)$ or $(-)$ correspond to $\varepsilon -\delta > 0$ and $\varepsilon -\delta < 0$, respectively. The last equation, in both cases, is the quadratic in $\varepsilon(\pi)$ implying the appearance of two bands, both for attractive and repulsive effective interaction.
		\begin{figure*}[t!]
			\subfigure[]{\includegraphics[width=0.35\textwidth]{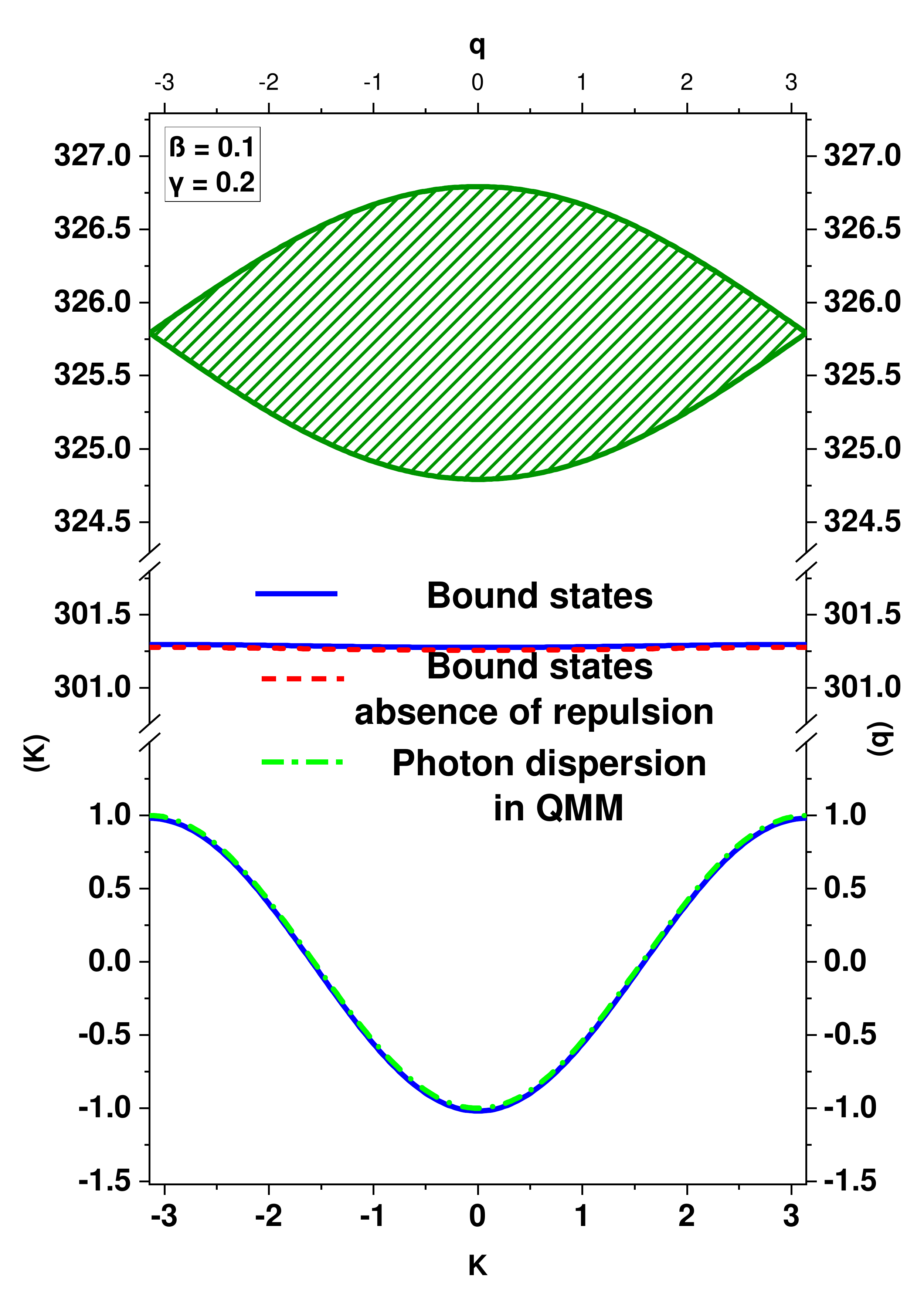}}
			\subfigure[]{\includegraphics[width=0.35\textwidth]{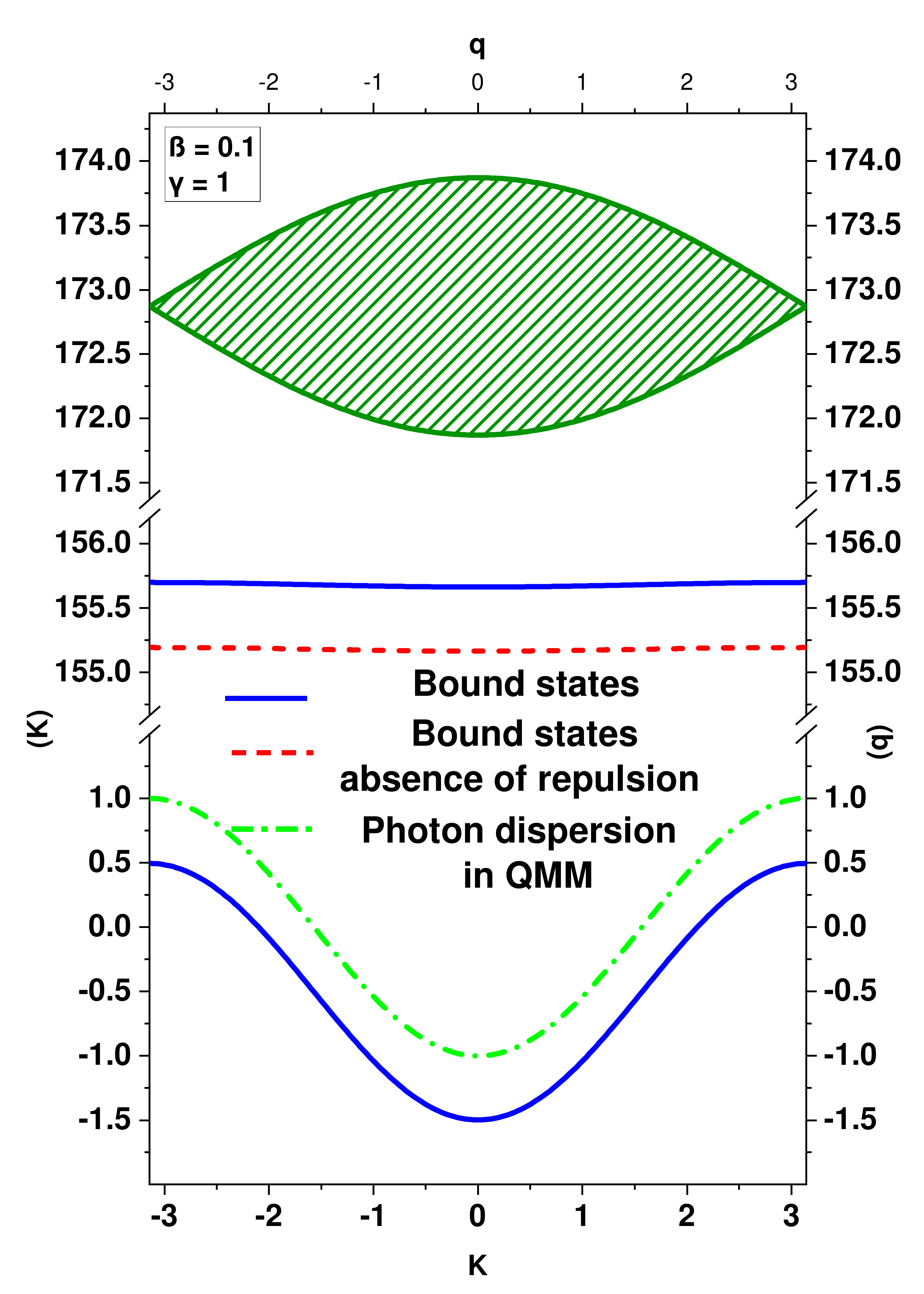}} 
			\subfigure[]{\includegraphics[width=0.35\textwidth]{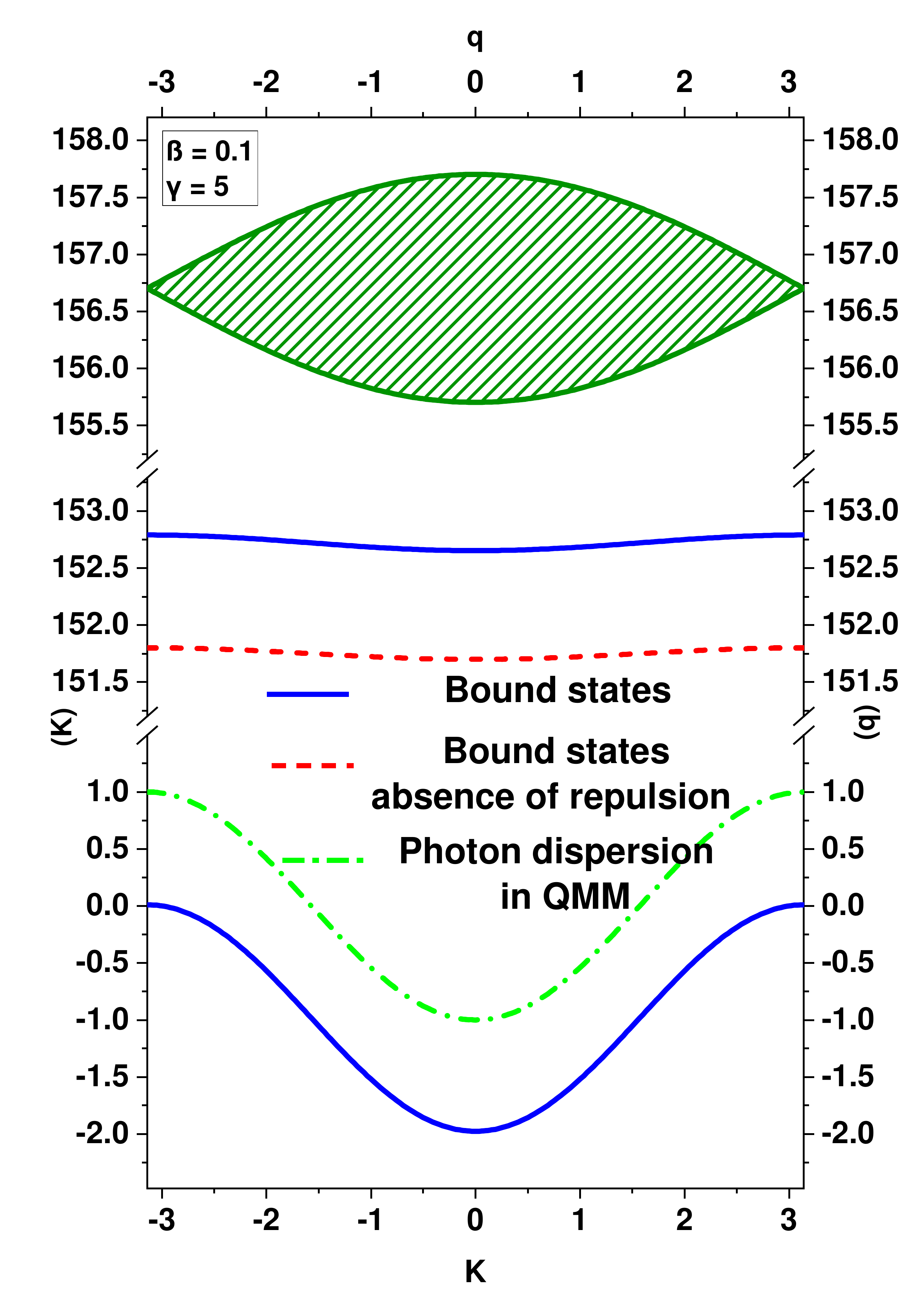}} 
			\subfigure[]{\includegraphics[width=0.35\textwidth]{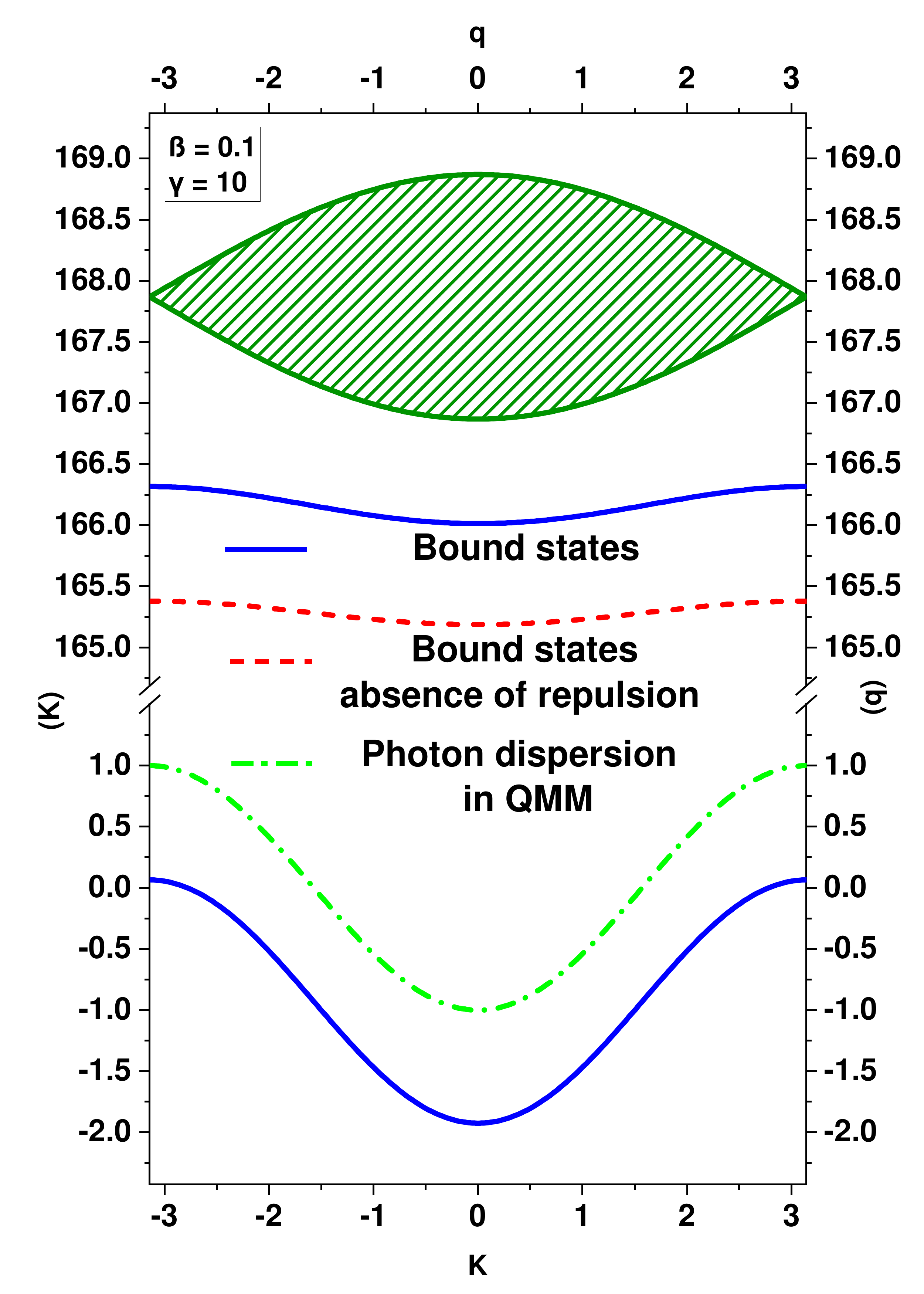}}
			\caption{Graphical illustration of the energy spectrum ($\varepsilon(K)$) of system for $\beta =0.1$, and for four different $\gamma$. Green shaded area corresponds to free states. Blue solid lines correspond to qubit--photon bound states. For the comparison we gave a band of bound states in the case in the absence of repulsive interaction--red dotted lines, and pure photon dispersion curve $\varepsilon(q)$ -- green dotted lines.}
			\label{fig02}
		\end{figure*}
		\begin{figure*}[t!]
			\centering
			\subfigure[]{\includegraphics[width=0.35\textwidth]{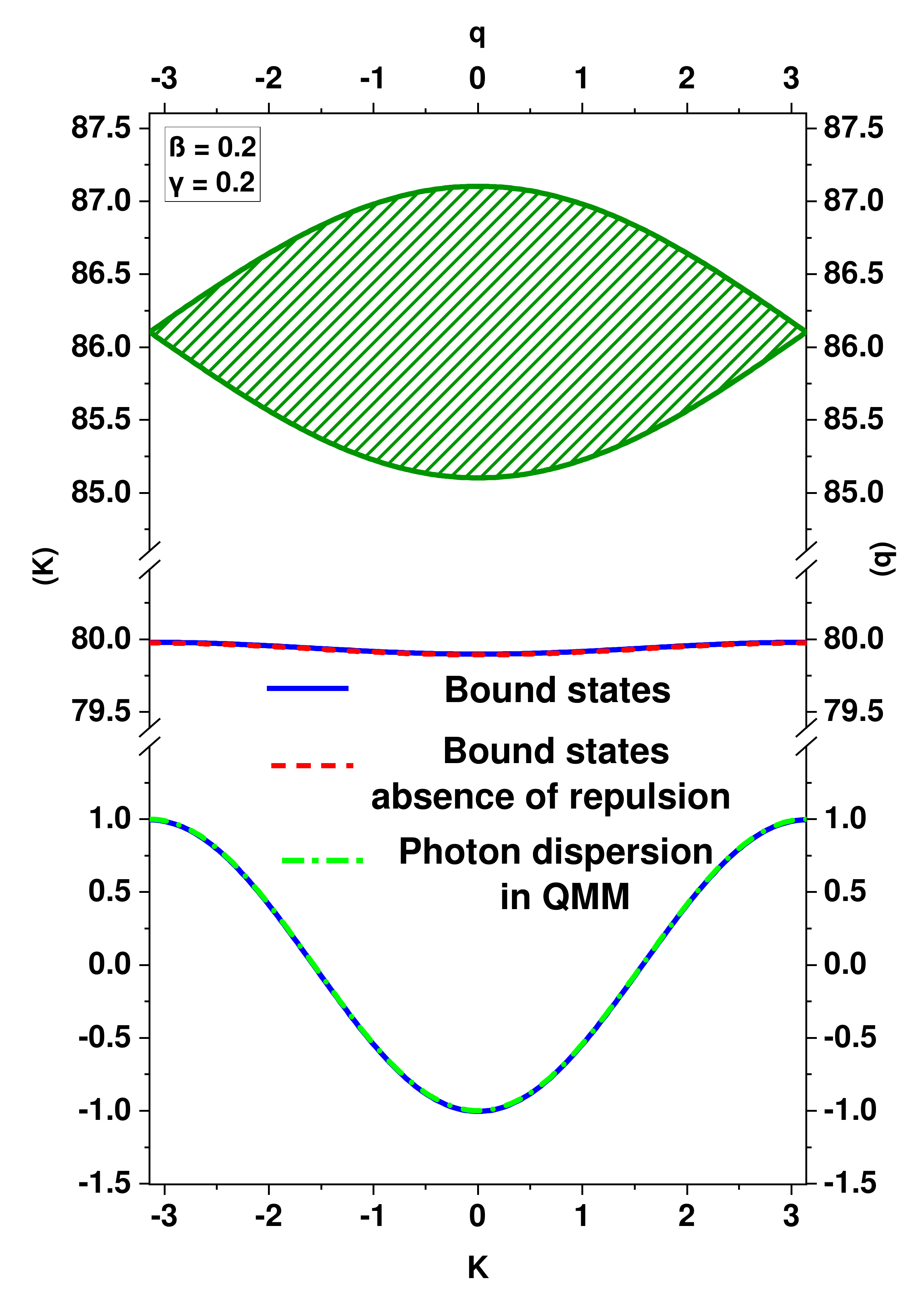}} 
			\subfigure[]{\includegraphics[width=0.375\textwidth]{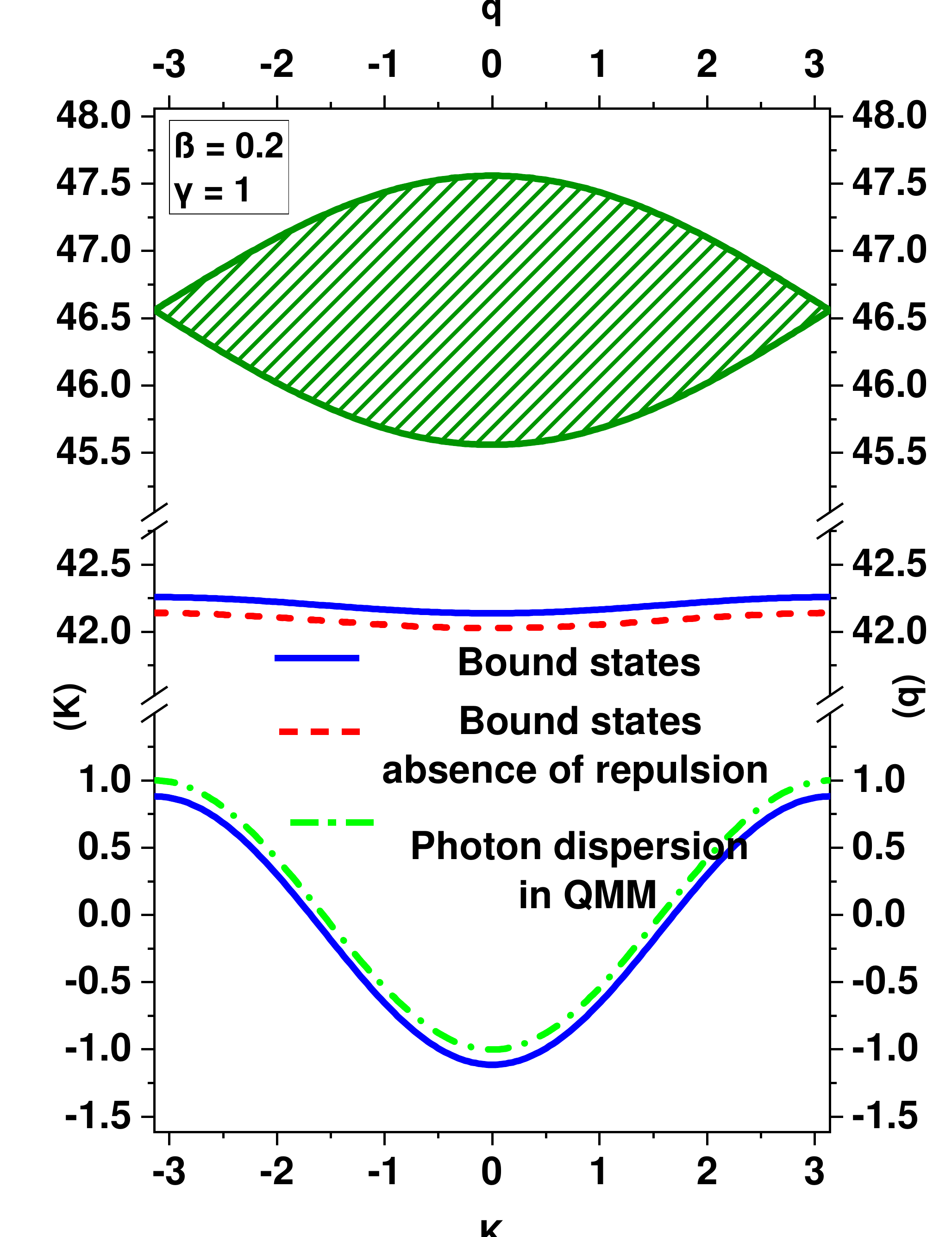}} 
			\subfigure[]{\includegraphics[width=0.35\textwidth]{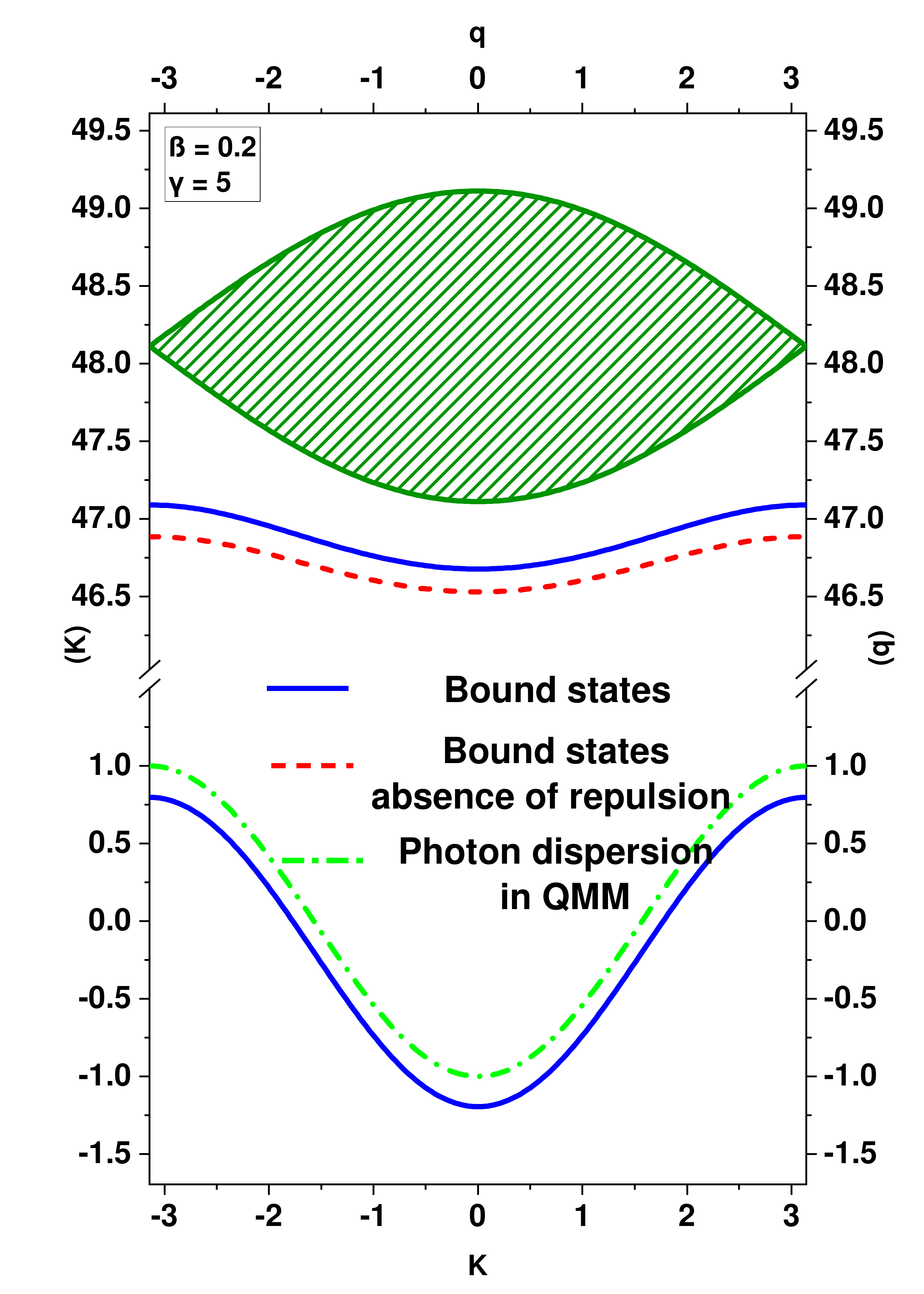}} 
			\subfigure[]{\includegraphics[width=0.35\textwidth]{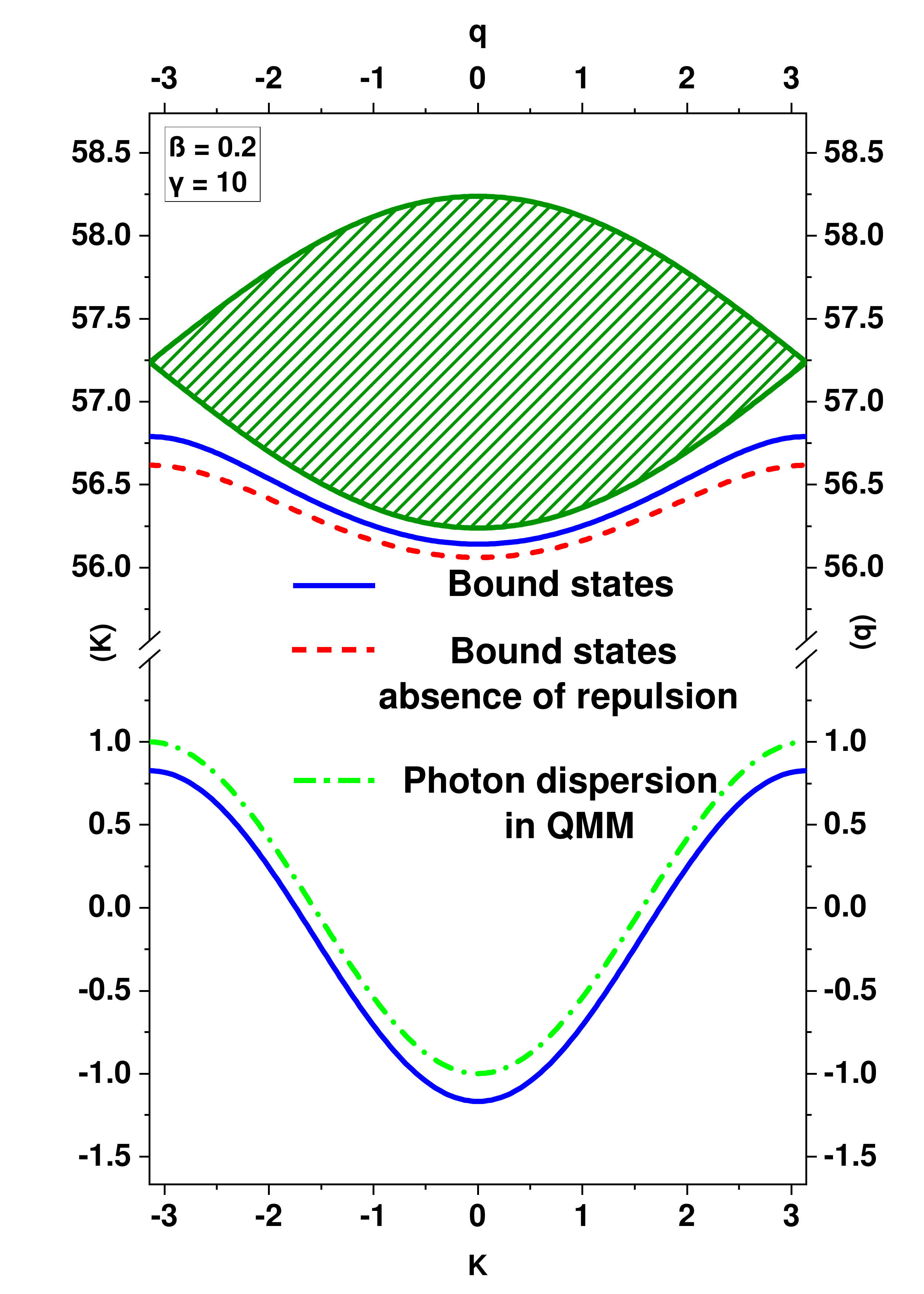}}
			\caption{ Energy spectrum ($\varepsilon(K)$) of system for $\beta =0.2$, and for the same values of $\gamma$ as in the preceding case.}
			\label{fig03}
		\end{figure*}
		\begin{figure*}[t!]
			\centering
			\subfigure[]{\includegraphics[width=0.35\textwidth]{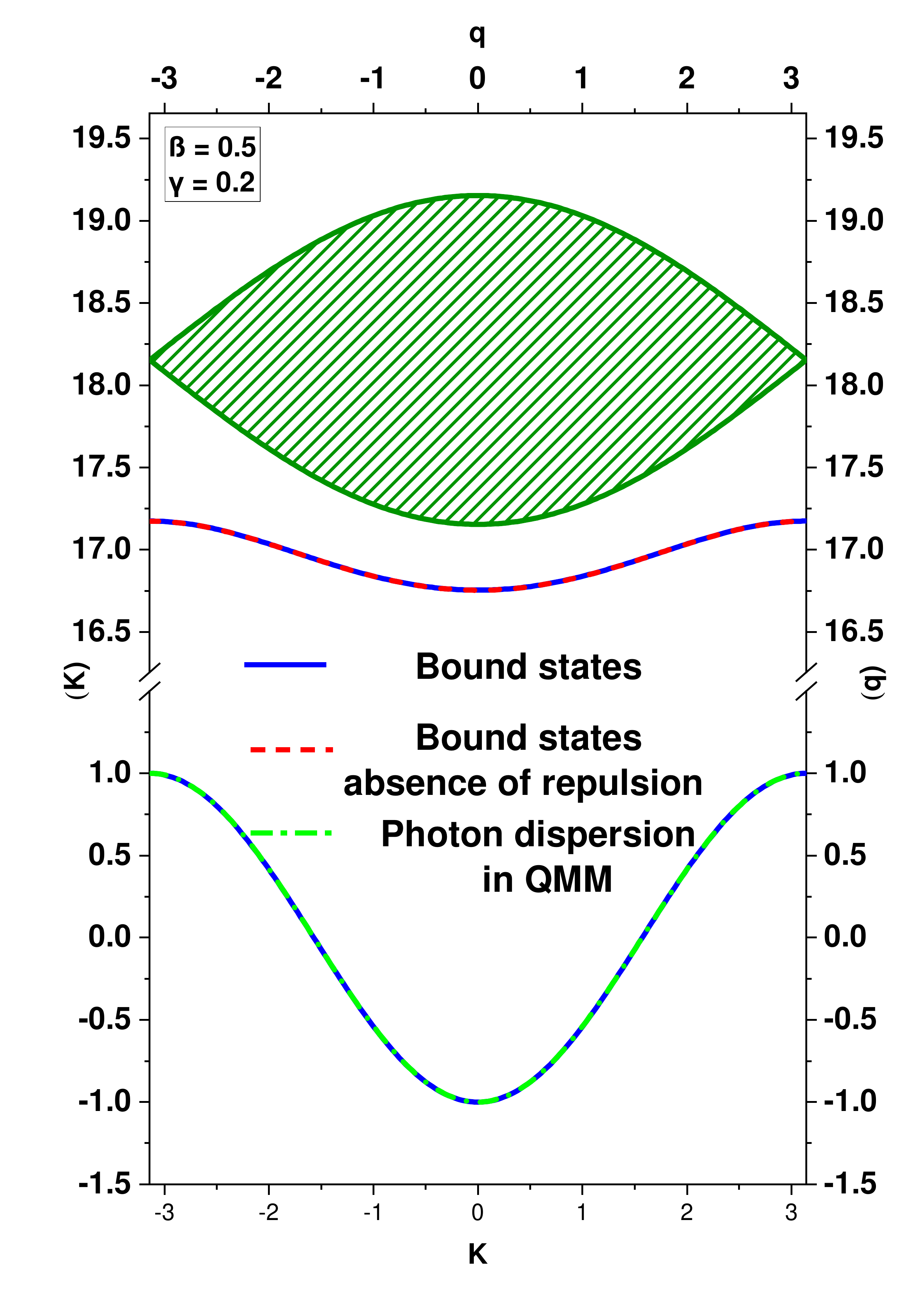}} 
			\subfigure[]{\includegraphics[width=0.35\textwidth]{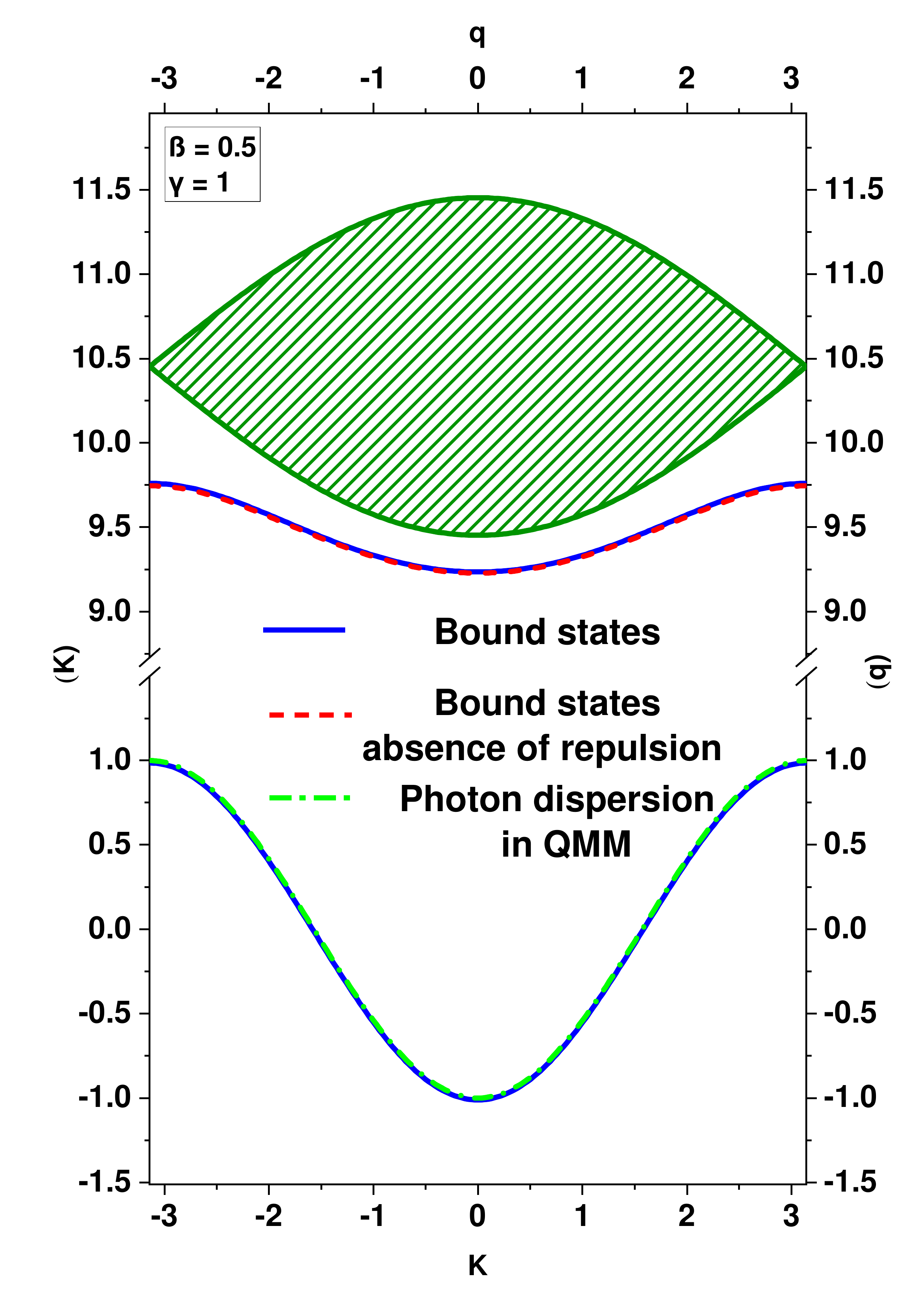}} 
			\subfigure[]{\includegraphics[width=0.35\textwidth]{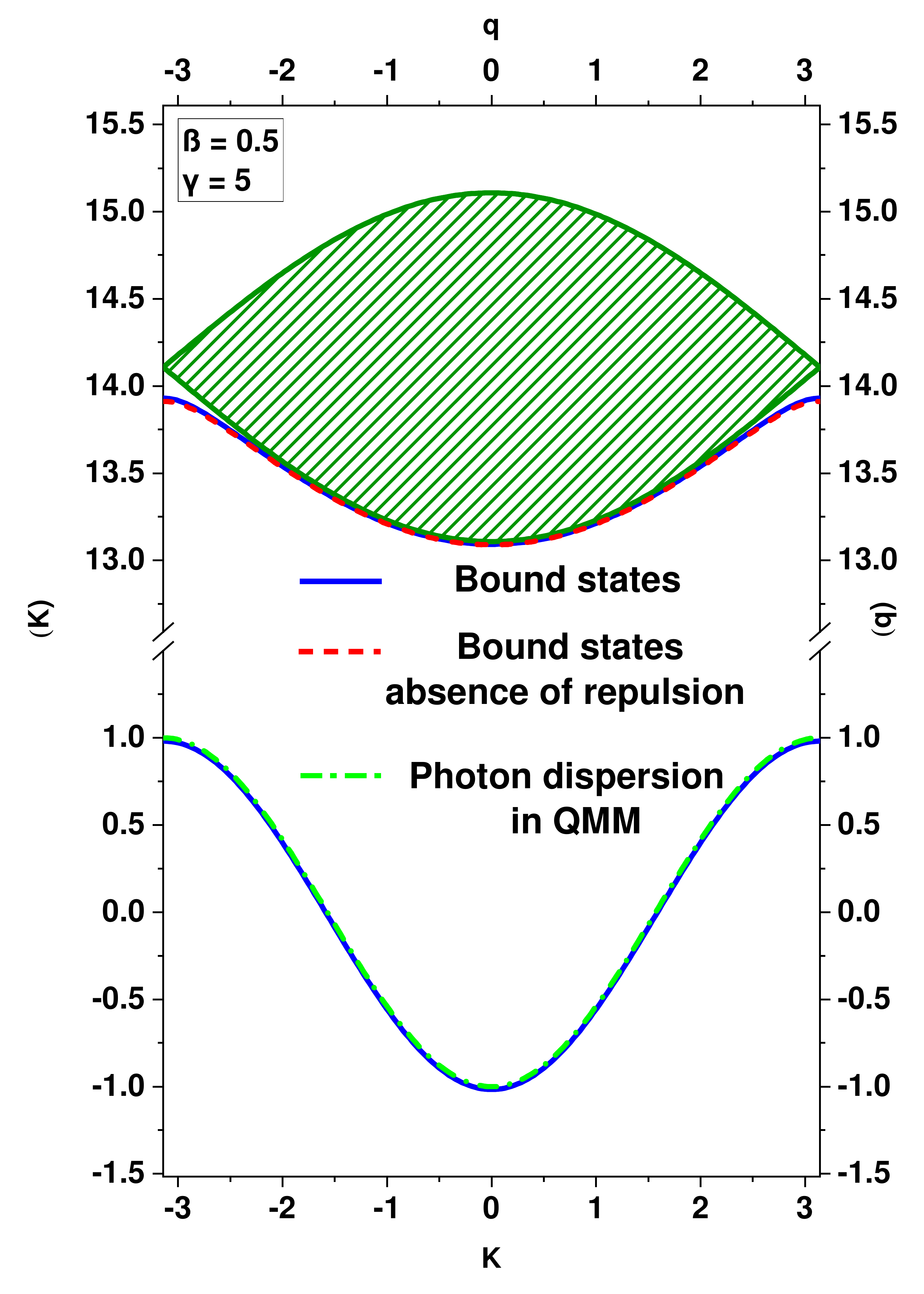}} 
			\subfigure[]{\includegraphics[width=0.35\textwidth]{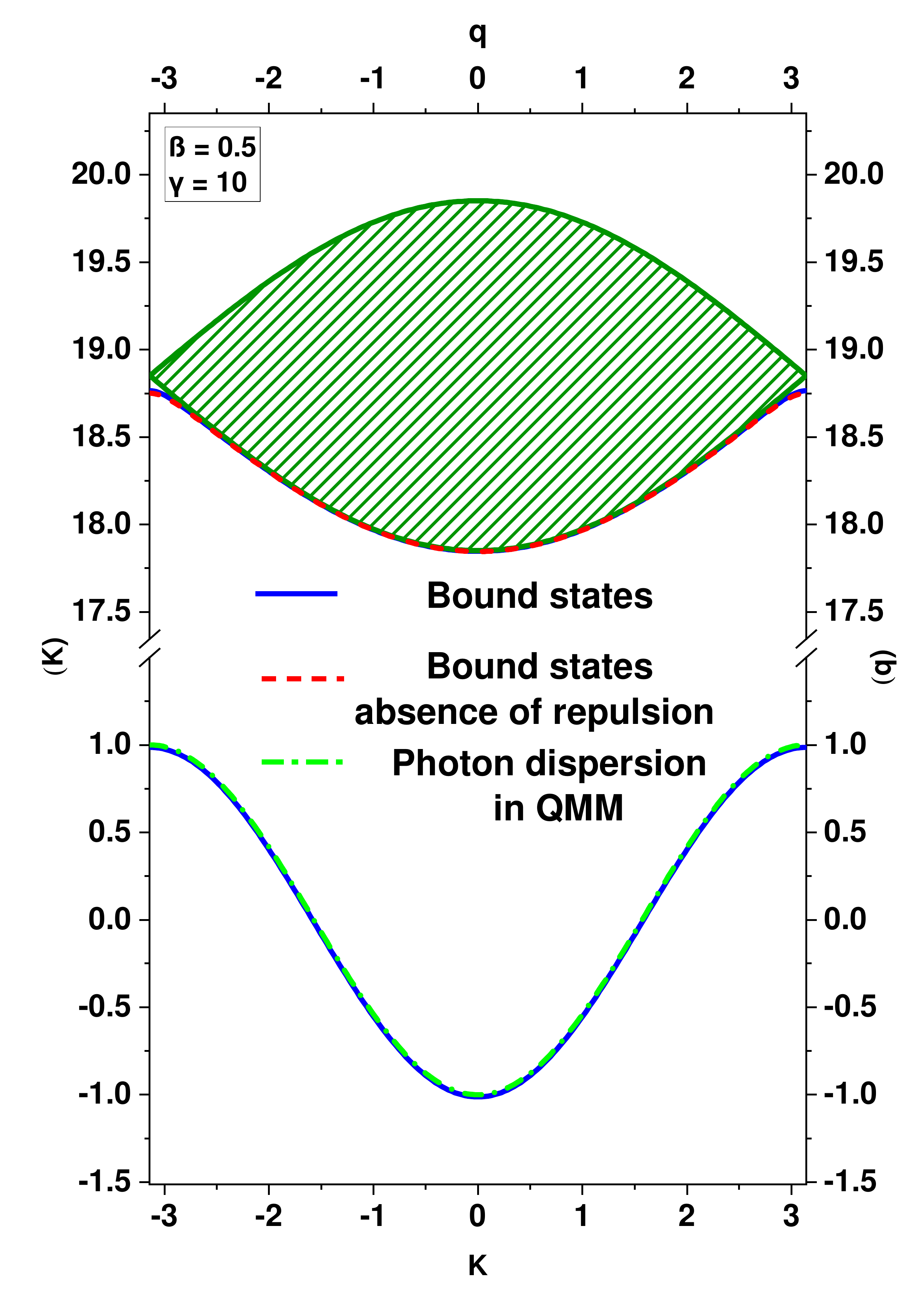}}
			\caption{ Same as in previous cases for $\beta=0.5$.}
			\label{fig04}
		\end{figure*}
		Solutions of Eq. (\ref{edge}) are: 
		\begin{eqnarray}
			\label{attr}
			\non \varepsilon_{\pm}(\pi)=&&\frac{1+\delta-a}{2}\pm\frac{1-\delta+a}{2}
			\sqrt{1+\left(\frac{a\gamma}{1-\delta+a} \right)^2 }.\\ 
			\non && \mathrm{for\;\; attractive\;\; effective \;\;interaction},\\
			\non \varepsilon_{\pm}(\pi)&&=\frac{1+\delta+a}{2}\pm\frac{1-\delta-a}{2}
			\sqrt{1-\left(\frac{a\gamma}{1-\delta-a} \right)^2 },\\
			&& \mathrm{for\;\; the\; repulsive\;\;one.}
		\end{eqnarray} 
		In the present context $\delta$ is large as compared with other system parameters. Thus, the ratios in both square roots may be regarded as small quantities. This enables us to expand both square roots in Eq. (\ref{attr}) in terms of "small parameter" $(a\gamma/(1-(\delta\pm a)))^2$ which yields the corresponding asymptotic relations:
		\begin{eqnarray}
			\label{asym}
			\non \varepsilon_{-}\approx&&\delta - a -\frac{(\frac{a\gamma}{2})^2}{1-\delta+a},\\
			\non \varepsilon_{+}\approx&& 1+\frac{(\frac{a\gamma}{2})^2}{1-\delta+a},\\
			\non&& \mathrm{For\; "attractive" \;effective\; interaction,}\\
			\non \varepsilon_{-}\approx&&\delta + a -\frac{(\frac{a\gamma}{2})^2}{\delta+a-1},\\
			\varepsilon_{+}\approx&& 1+\frac{(\frac{a\gamma}{2})^2}{1-\delta-a},\\&&
			\non \mathrm{for \; "repulsive"\; effective \; interaction.}
		\end{eqnarray}
		Based on these equations we may estimate under which conditions particular types of solutions exist. For that purpose, we recall the existence conditions of the solutions--\textit{subsection C}. We focus on repulsive interaction for which our numerical calculations do not find meaningful solutions for reliable parameter values. According to Eq. (\ref{eigen1}) its solutions exist provided that $a'(K)>0$. Substituting the corresponding asymptotic solution from Eq. (\ref{attr}), the third equation in Eq. (\ref{asym}), into $a'(\pi)$ we obtain the following condition:
		\begin{equation}
			\label{conit}
			a(\frac{\gamma}{2})^2>\varepsilon(\pi)-1\Leftrightarrow \delta+a<1+ 
			a(\frac{\gamma}{2})^2 \;\;\mathrm{for\; \; \varepsilon_+}.
		\end{equation} 
		On the other hand, solution $\varepsilon_-$, after subtracting the $\delta$ on both sides, attains the form: $$ \varepsilon -\delta=1-\delta+\frac{\frac{\gamma^2}{4}}{\delta+a-1}.$$ Note that neither of these conditions can be satisfied in the present case. Namely, the condition for the existence of solutions in the case of repulsive interaction reads $\varepsilon-\delta<0$, which cannot be satisfied in practice due to large values of $\delta$. In particular, for that purpose $\gamma \gtrsim 100$ is required.
		\subsection{Solutions: numerical results}
		Numerical calculations, were performed for the values of system parameters covering both \textit{charging (large $\gamma$)} and \textit{Josephson ( small $\gamma$)} regime. Note that there is no any particular restrictions on the value of dimensionless speed of light $\beta$ in QMM. In particular, in literature \cite{Rakhmanov2008,Shvetsov2013,Asai2018,scir} $\beta$ was taken to vary from few tenths up to \emph{1}. Here we restrict ourselves to $\beta \eqslantless 0.5$ since the results for its larger values do not exhibit any \emph{substantial qualitative} difference. Thus we used $\beta =0.1; 0.2,\; \mathrm{and}\; 0.5$, while, for each $\beta$, we took 
		four values $\gamma: 0.2,\; 1, \; 5\; \mathrm{and}\; 10$.
		Our results are illustrated in Figs. (\ref{fig02}) -- (\ref{fig04}). The energy spectrum consists of the free state continuum, green shaded area, and two bands of qubit photon bound states which are observed for each set of system parameters. The higher energy band (Band 1) lies below the free state continuum and, for large values of $\beta$ ($\beta= 0.5$ as presented at Fig.(4)), is practically indistinguishable from the bound states appearing in the case of pure attractive interaction corresponding to \textit{ad hoc} choice $B=0$. For small $\gamma$ Band 1 is well separated from the continuum approaching it for larger values of $\gamma$. Band 1 features profoundly change as $\beta$ decreases. For example, for $\beta=0.1$ (Fig. 3) the magnitudes of the Band 1 bound states energies and those of the free states, for each $K$, are almost twenty times higher than for $\beta=0.5$. In addition, for small $\gamma$ ($\gamma = 0.2$), Band 1 is practically indistinguishable from bound states corresponding to pure attractive interaction. As $\gamma$ rises Band 1 and solutions for the pure attractive interaction separate and both gradually tend towards the free state continuum. Qualitatively the same behavior is observed for $\beta = 0.2$ with a somewhat different degree of changes. 
		As presented at the lower part of figures (\ref{fig02})--(\ref{fig04}), in parallel with Band 1 the second one (Band 2), appears. This is the novel band lying deeply below Band 1. It emerges from the competition between the attractive and repulsive interaction and lies below the free photon band. Its dependence on parameters $\beta$ and $\gamma$ exhibits similar behavior as for Band 1. That is, for large $\beta$, irrespective of the values of $\gamma$, Band 2 and free photon band are practically identical, due to complete compensation of the effective attractive and repulsive interactions. That is QMM is fully transparent, and there are no bound states. For smaller values of $\beta$ attractive interaction dominate over the repulsive and qubit--photon bound states to emerge, providing that $\gamma$ is high enough. Nevertheless, QMM is still transparent but for qubit--photon bound states.
		\section{Concluding Remarks}
		In this paper, we have studied the energy structure and cooperative qubit--photon excitation of a one--dimensional superconducting quantum metamaterial. The system consists of the large number ($\mathcal{N}\gg 1$) periodically arranged charge qubits placed inside the massive two-strip superconducting resonator. In such a setup each unit cell of SCQMM (sketched at Fig.1 b) can be viewed as an electromagnetic resonator, while the system as a whole, represents a coupled-resonator (cavities) waveguide with single an "atom" per cavity. 
		This setup, upon quantization, exhibits some novel features in comparison to those used so far in the studies of the matter--light interaction. In particular, the system is translationally invariant since the number of "cavities" and "atoms" match: each cavity contains a single qubit. So far the studies on the subject were carried under the condition that the individual "atoms" \cite{aphbs3, aphbs4, aphbs5, aphbs6, aphbs7, prohib, aphbs8, aphbs9, aphbs10, aphbs11} or their ensembles \cite{aphbs0} are placed in different resonators and where translational invariance has been rarely accounted \cite{aphbs11, cohout}.
		Furthermore, the qubit--photon interaction is substantially different from that utilized in most studies on the subject \cite{aphbs3, aphbs4, aphbs5, aphbs6, aphbs7, aphbs0, aphbs8, prohib, aphbs9, aphbs10} which were carried out within the certain modifications of the celebrating Dicke model \cite{dicke}. The essential difference is that it now has two components: the attractive and the repulsive one originating on account of different mechanisms: i) simultaneous excitation ($\sigma^{\dg}a$) and deexcitation ($\sigma^{-}a^{\dg}$) of the $n$--th qubit by absorption and emission of the single-photon, attractive one and ii) the repulsive one from the photon scattering by qubits accompanied by their excitation and deexcitation.
		The main consequence of these peculiarities is the emergence of the mixed qubit--photon bound states. In particular, the energy spectrum of the qubit--photon bound states consists of two widely separated bands. The higher energy one lies far over the photon continuum. It is very close to that observed in the simple case of pure attractive interaction and appears for large $\varepsilon$ when $a'\rightarrow a$. The results, almost identical to the preceding ones \cite{bs1,bs2}, were observed. The lower band, near the band edges, lies within the photon continuum. 
		Based on the recent findings \cite{aphbs8,aphbs9} we expect that these bound states may exert a considerable influence on the photon transport properties. It relies upon the possibility of radiation trapping due to the creation of these bound states \cite{aphbs8,aphbs9}. In the present case, due to translational invariance of the system, radiation trapping concerns the qubit dressing by photon cloud. The formation of bands of such complexes implies their free propagation. Band flattening with changes of the values of system parameters points to the slowing down of these mixed states. The emergence of the flat bands implies a possible stopping light which indicates that the proposed setup could be used for manipulating the opens up the novel means for realizing operable quantum devices. 
		The proposed setup is convenient for the practical realization of such devices with controllable parameters which could be achieved by applying a constant external magnetic field in parallel with propagating EM field. In such a way, vector potential attains an additional constant term $\alpha_n\rightarrow \alpha_n+\alpha_0$ so that interaction term in (\ref{tot}), after straightforward calculation, reads:
		$$ H_i\approx-2E_J\cos\varphi_n\left[ \cos\alpha_0\left( 1-\frac{\alpha^2_n}{2}\right) -\sin\alpha_0 \;\alpha_n\right]. $$
		Varying external field it would possible to change the tunneling energy and to "flip" between different regimes. A particularly interesting situation arises when $\alpha_0=\pi/2$ when the interaction term, upon quantization, attains the form identical to that encountered in coplanar arrangements. 
		Finally, let us comment on the generality of our results. We do not expect that the features of the propagating signal, in the proposed geometrical arrangement, should not qualitatively depend on the particular choice of the type of qubit \cite{zag}. Thus, for the simplicity and certain flexibility for the manipulation of the single qubit, we use here charge qubits, while any other type would give analogous results. 
		\begin{acknowledgments}
			We thank D. Kapor for fruitful discussion and useful comments on the manuscript. 
			This work was supported by the Ministry of Education, Science, and
			technological development of the Republic of Serbia. Z.I. acknowledges support by the "Vin\v ca" Institute -- special grant No. 104-1-2/2020-020, dated 11.01.2021. We also acknowledge the co-financing of this research by the European Union and Greek national funds through the Operational Program Crete 2020-2024, under the call "Partnerships of Companies with Institutions for Research and Transfer of Knowledge in the Thematic Priorities of RIS3Crete", with project title "Analyzing urban dynamics through monitoring the city magnetic environment" (project KPHP1 - 0029067) and also by the Ministry of Science and Higher Education of the Russian Federation in the framework of Increase Competitiveness Program of NUST "MISiS" (No. K2-2019-010), implemented by a governmental decree dated 16th of March 2013, N 211. N.L. acknowledges support by the General Secretariat for Research and Technology (GSRT) and the Hellenic Foundation for Research and Innovation (HFRI) (Code No. 203).
		\end{acknowledgments}
		\section{Appendix 1: quantization of the model Hamiltonian}
		\subsection{Quantization of the qubit subsystem}
		After expansion $\cos\alpha_n\approx 1-\alpha^2_n/2$, and transition in Cooper pair basis number basis $|N\rangle$ together with the correspondence: $\hat N=-i\frac{\partial}{\partial \phi_n}$ and noticing that $e^{\pm i\hat \varphi_n }|N\rangle = |N \pm 1\rangle$ we rewrite Hamiltonian Eq. (\ref{tot}) in the charge basis as follows:
		\begin{eqnarray}\label{CPN}
			\non H=&&\sum_n 2E_C\hat N^2_n|N\rangle_n\langle N|-\\ \non && E_J\sum_n|N\rangle_n\langle N+1|+|N+1\rangle_n\langle N|+\\ \non && \frac{E_J}{2}\sum_n\bigg(|N\rangle_n\langle N+1|+|N+1\rangle_n\langle N|\bigg)\alpha^2_n+ \\ &&
			\sum_n\left(\frac{2\hbar^2}{E_c}\dot\alpha^2_n+ E_{em}(\alpha_{n+1}-\alpha_n)^2\right)
		\end{eqnarray} 
		In the reduced state space, in which the single island can be unoccupied ($N=0$) or occupied by a single Cooper pair ($N=1$) we obtain reduced Hamiltonian:
		\begin{eqnarray}\label{reduced} \non&&
			H=-E_c\mathcal{N}+ \sum_n\bigg[E_c \tau^z_n-E_J\tau^x_n\bigg]+\\
			&&\sum_n\left(\frac{2\hbar^2}{E_c}\dot\alpha^2_n+ E_{em}(\alpha_{n+1}-\alpha_n)^2+\frac{E_J}{2}\tau^x_n\alpha^2_n\right).
		\end{eqnarray}
		where $\tau^x_n=|1\rangle_n\langle 0|+|0\rangle_n\langle 1|$ and $\tau^z_n=|1\rangle_n\langle 1|-|0\rangle_n\langle 0|$, while in deriving the above result we have used an apparent relation $\hat N_n=|1\rangle_n\langle 1|+|0\rangle_n\langle 0|\equiv 1$
		Qubit component of this Hamiltonian may be diagonalized by means of the norm preserving ( $1=|e\rangle_n\langle e|+|g\rangle_n\langle g|$) transformation:
		\begin{eqnarray}\label{UT}
			\non
			\tau^x_n=\cos\eta\left( |e\rangle_n\langle g|+|g\rangle_n\langle e|\right)-\sin\eta \left( |e\rangle_n\langle e|-|g\rangle_n\langle e|\right),\\
			\tau^z_n=\cos\eta\left( |e\rangle_n\langle e|-|g\rangle_n\langle g|\right)+\sin\eta \left(|e\rangle_n\langle g|+|g\rangle_n\langle e|\right),\\
			\non \tan\eta=\frac{E_J}{E_C},\;\sin\eta=-\frac{E_J}{\sqrt{E^2_c+E^2_J}},\; \cos\eta=\frac{E_C}{\sqrt{E^2_c+E^2_J}}
		\end{eqnarray}
		In such a way, up to an irrelevant constant, above Hamiltonian became
		\begin{widetext}
			\begin{eqnarray}\label{htl}
				\non H= &&\sum_n \left\{ 2\epsilon|e\rangle_n\langle e|+\left[\frac{E_JE_c}{2\epsilon}\left(|e\rangle_n\langle g|+|g\rangle_n\langle e|\right)-\frac{E^2_J}{\epsilon}|e\rangle_n\langle e|\right]\alpha^2_n \right\}+ \\ && \sum_n\left(\frac{2\hbar^2}{E_c}\dot\alpha^2_n+ E_{em}(\alpha_{n+1}-\alpha_n)^2+
				\frac{E^2_J}{2\epsilon}\alpha^2_n\right).
			\end{eqnarray}
		\end{widetext}
		Here $\epsilon=\sqrt{E^2_c+E^2_J}$, so that the $\pm \epsilon$ denote the energy eigenstates: ground (-) and excited (+) one. 
		\subsection{Quantization of EM field}
		As usual we consider $\alpha_n \ll 1$ and expand corresponding \textit{"cosine"} term in interaction. First we define the generalized momentum $P_n=\frac{2\hbar^2}{E_c}\dot{\alpha_n}$ canonically conjugated to $\alpha_n$. Now we treat photon variables as operators $\alpha_n\rightarrow\hat \alpha_n$, $P_n\rightarrow \hat P_n$ requiring that they satisfy commutation relation: $[\alpha_n, P_m]=i\hbar\delta_{m,n}$, which holds for the following transformation:
		\begin{eqnarray}
			\label{quantiz2}
			\non \hat \alpha_n=\frac{1}{2}\sqrt{\frac{E_C}{\hbar\omega}}\left(a_n+a^{\dg}_{n}\right),\;\;
			\hat P_n={i\hbar}\sqrt{\frac{\hbar\omega}{E_c}}\left(a^{\dg}_{n}-a_n\right),
		\end{eqnarray}
		Substitution of the above expressions in Eq. (\ref{htl}) yields the following model Hamiltonian:
		\begin{widetext}
			\begin{eqnarray}\label{hf}
				\non H=\non\sum_n \left[2\epsilon|e\rangle_n\langle e|+\hbar\omega a^{\dg}_n a_n-\frac{E_{em}E_C}{2\hbar\omega}a^{\dg}_n\left(a_{n+1}+a_{n-1}\right) \right] 
				+\frac{E_JE_C}{8\hbar\omega \epsilon}\sum_n\left[ E_c\left(|e\rangle_n\langle g|+|g\rangle_n\langle e|\right)-2E_J|e\rangle_n\langle e|\right]\left(a^{\dg}_n+a_n \right)^2
			\end{eqnarray}
		\end{widetext}
		\begin{widetext}
			\begin{eqnarray}\label{modH}
				\non H_s=\Delta\sum_n|e\rangle_n\langle e|+\hbar\omega\sum_n a^{\dg}_n a_n-J\sum_n a^{\dg}(a_{n+1}+a_{n-1})-\sum_n\left[A|e\rangle_n\langle e|-B(|e\rangle_n\langle g|+|g\rangle_n\langle e|) a^{\dg}_n a_n\right]
			\end{eqnarray}
		\end{widetext}
		\begin{equation}
			\hbar\omega=\sqrt{2E_{em}E_C+\frac{E_CE^2_J}{2E}},\;\; 
			J=\frac{E_{em}E_C}{2\hbar\omega},\;\;
			A=\frac{E^2_JE_c}{4\hbar\omega E},\;\; 
			B=\frac{E_JE^2_c}{8\hbar\omega E}.
		\end{equation}
		
	\end{document}